\documentclass[article,aps,prl,twocolumn,groupedaddress]{revtex4}
\usepackage{inputenc}
\usepackage{braket}
\usepackage[rightcaption]{sidecap}
\usepackage{dcolumn}
\usepackage{bm}% bold math
\usepackage{amsmath,amssymb,amsthm,euscript,graphicx}
\makeatletter
\def\maketag@@@#1{\hbox{\m@th\normalfont\normalsize#1}}
\usepackage{graphicx,color}
\usepackage{verbatim}
\graphicspath{{./Figs/}}
\usepackage{tikz}
\usetikzlibrary{arrows.meta, positioning}
\tikzset{
	pics/AxisRotator/.style={
		code={
			\draw [x=1em, y=1em, line width=.2ex, -{Latex[length=.5em, quick]},rotate=#1] (-.25,-.7) arc (-150:165:.3375 and 1.375);
	}},
	pics/AxisRotator/.default=0
}
\usepackage{pgfplots}

\pgfplotsset{every axis/.append style={
		axis x line=middle,    % put the x axis in the middle
		axis y line=middle,    % put the y axis in the middle
		axis line style={<->}, % arrows on the axis
		xlabel={$x$},          % default put x on x-axis
		ylabel={$y$},          % default put y on y-axis
}}

\tikzset{>=stealth}
\usepackage{subfigure}
\usepackage{epstopdf}
\usepackage{color}
\usepackage{natbib}
\usepackage[strict]{changepage}
\usepackage{epsfig}
\usepackage{fancybox}
\usepackage{listings}
\usepackage{gensymb}
\usepackage{url}
\usepackage{verbatim}
\usepackage{float}
\usepackage{pgfplots}
\usepackage{mathrsfs}
\usepackage[title]{appendix}
\newcommand{\eval}[2][\right]{\relax \ifx#1\right\relax \left.\fi#2#1\rvert}

\def\be{\begin{equation}}
\def\ee{\end{equation}}
\def\bes{\begin{eqnarray}}
\def\ees{\end{eqnarray}}
\def\2{\frac{1}{2}}
\def\4{\frac{1}{4}}

\begin{document}
	
%	\noindent\fbox{%
%		\parbox{\textwidth}{%
%			This manuscript has been authored by UT-Battelle, LLC under Contract No. DE-AC05-00OR22725 
%			with the U.S. Department of Energy. The United States Government retains and the publisher, 
%			by accepting the article for publication, acknowledges that the United States Government retains 
%			a non-exclusive, paid-up, irrevocable, world-wide license to publish or reproduce the published 
%			form of this manuscript, or allow others to do so, for United States Government purposes. The 
%			Department of Energy will provide public access to these results of federally sponsored research
%			in accordance with the DOE Public Access Plan (http://energy.gov/downloads/doe-public-access-plan).
%		}%
%	}
%	\clearpage
	
	% !TeX spellcheck =  en_GB
	%\preprint{APS/123-QED}
	%\title{Quantum plasmonics in hyperboloidal  and paraboloidal domains}
	\title{{General Approach to Study Geometric Effects on Classical \& Quantum Fields \& Eigenmodes of Particles with Finite \& Infinite Extend for Applications in Plasmonics \& Scanning Probe Microscopy}}
	\author{M. Bagherian$^1$}\email{mbagherian@mail.usf.edu}
	\affiliation{$^1$Department of Mathematics and Statistics, University of South Florida, Tampa, Florida 33620}
	\date{\today}
	\begin{abstract}
		This manuscript provides a general approach to the investigation of field quantization in high-curvature geometries. The models and calculations can help with understanding the elastic and inelastic scattering of photons and electrons in  nanostructures and probe-like metallic domains. The results find important applications in high-resolution photonic and electronic modalities of scanning probe microscopy, nano-optics,  plasmonics, and quantum sensing. 
		Quasistatic formulation, leading to nonretarded quantities, is employed and justified on the basis of the nanoscale, here subwavelength, dimensions of the considered domains of interest.
		Within the quasistatic framework,  the nanostructure material domains with  frequency-dependent dielectric functions is presented. Quantities associated with the normal modes of the electronic systems, the nonretarded plasmon dispersion relations, eigenmodes, and  fields are then calculated for several geometric entities of use in nanoscience and nanotechnology. 
		From the classical energy of the charge density oscillations in the modeled nanoparticle, the Hamiltonian of the system, which is used for quantization, is derived.
		The quantized plasmon field is obtained and, employing an interaction Hamiltonian derived from the first-order perturbation theory within the hydrodynamic model of an electron gas,  an analytical expression for the radiative decay rate of the plasmons could be obtained. 
		The established treatment could be applied to multiple geometries  to investigate the quantized charge density oscillations on their bounding surfaces. For more on this, author is reffred to \cite{Bag2018, garapati2,meDiss}.
	\end{abstract}
	%\pacs{47.45.Nd, 68.37.Ps, 47.45.Gx, 47.45.Dt}                        
	\maketitle
	%%%%

\section{Introduction}\label{into}

Interaction of photos with surface plasmons has been studied extensively during the past five decades, particularly in relation to their application in scanning probe microscopy (SPM), they present a high potential for emerging applications in fields such as quantum sensing~\cite{garapati2, ben,qafm,tlf:thin}. It is known as a branch of microscopy that forms images of surfaces using a physical probe. Throughout these studies, various geometries have been investigated. In particular, Ritchie together with Crowell, Little, Ashley and Ferell \cite{Crowell, Little_Paper, B, RIT34,Ritchie} have studied the interaction of general and special cases of surface of a metallic sphere and oblate spheroid with photons and derives the very first relations describing the field quntization of surface plasmons. Inspired by these results, quantization relations for a special case of a long string-like cylinder has been studied by Burmistrova \cite{BURMISTROVA}. In this work, we wish to investigate the interaction in more detail as well as to extend the obtained results to both new finite and infinite geometries such as cylinder, paraboloid, hyperboloid, prolate spheroid and  the non-simply connected case of a torus. It should be pointed out that the \emph{surface plasmon-photon interaction} (SPP)  can be analyzed for various cases such as absorption, emission, Thomson and Rayleigh scattering, and  scattering cross-sections. Our main focus in this study, however, is the emission case and its application to the decay rate.  As we shall see later in this chapter, there is a close relation between both cases of emission and absorption in terms of the interaction matrix element. As a result, our study also addresses the absorption phenomena \cite{B,GENZEL}.

Throughout this dissertation, we consider a metallic particle confined in vacuum whose shape is described by one of the above mentioned geometries. The conduction electrons and ion centers in a metal together to form a plasma. The authors in \cite{pines}, while studying the collective oscillations in the metallic plasma, showed that through the plasma system exhibits resonances along the directions that charge density waves propagate, with no effect of damping. Ritchie \cite{Ritchie} in 1957 showed that the electron density can be supported at the surface of a bounded plane free-electron gas. In \cite{Ferrell1, SternFerrell}, it has been discussed both the property of plasmons and their interaction with charged particles. The properties of plasmons and details on how they interact with light or charges particles, such as photons, are also discussed and reviewed in \cite{FIS,MASRI, BARKER, RIT34,raether}. To describe the properties of a metal, mostly from a dynamical point of view, the complex-valued \emph{dielectric function} $\varepsilon(\omega)$ was used. Generally speaking, the dielectric function specifies the relation between the frequency and the wave-vector in a certain metal. It is also known as \emph{permitivity}. The main assumption here is that in the incident of large wave-vectors, the dielectric function would be no longer wave-vector-dependent \cite{pines}. 

Throughout this chapter, we adopt the conventions of Ritchie mostly presented in \cite{Little_Paper}, unless otherwise specified. 
The \emph{dielectric function}, $\varepsilon(\omega)$, is mostly used to express certain properties of a metal. It is a complex and frequency-dependent function which in general also depends on wave vector of the plasmon field, $k_p \equiv 2\pi/ \lambda_p$ where $\lambda_p$ denotes the plasmon wavelength. However, under certain suitable assumptions, the wave vector dependency could be ignored \cite{pines}. Under the assumption that the wavelength is large enough, the dielectric function can be treated as independent from the wave vector. This requirement is met for all cases which have been investigated in this work.\\

\section{Free-electron-gas dispersion relation}

We start by considering the case of a spring oscillator, with no damping forces. The equation of motion, using \emph{Newton's second law}, can be written as \cite{may}:
\begin{equation}\label{may1}
F(\mathbf r_i)=m\, \frac{d^2 \mathbf r_i}{dt^2} , 
\end{equation}
where $m$ denotes the mass and $\mathbf r_i$ is the displacement vector of $i$-th electron with respect to the equilibrium position. \emph{Hooke's} law, on the other hand, gives:
\begin{equation}\label{may2}
F(\mathbf r_i)=-k\, \mathbf r , 
\end{equation}
where $k$ denotes the spring constant. From Eqs.~\eqref{may1} and \eqref{may2}, one could write the second order differential equation: 
\begin{equation}\label{may3}
m\, \frac{d^2 \mathbf r_i}{dt^2} =-k\, \mathbf r_i, 
\end{equation}
whose solutions are given by:
\begin{equation}\label{may4}
\mathbf r_i(t)= A\cos(\omega_0t)+ B\sin(\omega_0t), 
\end{equation}
for some constants $A$ and $B$, where $\omega_0 = \sqrt{\frac{k}{m}}$ is the \textit{resonant frequency}. The equation of motion for the $i$-th electron, including both the driving forces and the \emph{damping} $\gamma$ is written as \cite{fox}:
\begin{equation}\label{may5}
\frac{d^2 \mathbf r_i}{dt^2}+ \gamma \frac{d \mathbf r_i}{dt}+ \omega^2_0 \mathbf r_i =- \frac{e}{m_e}\, \vec E, 
\end{equation}
where $e$, $m_e$ denote the electric charge and the mass of an electron, respectively, and $\vec E$ represents the \emph{electric field}. The terms $\omega^2_0 \mathbf r_i$ and $- \frac{e}{m_0}\, \vec E$ represent the \emph{restoring} and electric forces, respectively. The time dependent electric field of a light wave inducing oscillations is given by: 
\begin{equation}\label{may6}
\vec E(t)= \cos(\omega t+ \theta), 
\end{equation}
where $\omega$ is the frequency of the system and $\theta$ is denotes the phase of the wave. Eq.~\eqref{may6} could be written as:
\begin{equation}\label{may7}
\vec E(t)=\vec E_0 \, \text {Re}\, ({e^{-(i\omega t+\theta)}}), 
\end{equation}
where $\vec E_0$ denotes the amplitude. We are mostly interested in those solutions of Eq.~\eqref{may5} which have the form similar to Eq.~\eqref{may7}, namely, 
\begin{equation}\label{may8}
\mathbf r_i(t)=\vec R_0 \,  \text{Re} ({e^{-(i\omega t+\tilde \theta)}}),  
\end{equation}
where $ \vec R_0$ and $ \tilde \theta$ represent the amplitude and phase of the oscillations, respectively. Substituting Eqs. ~\eqref{may7} and \eqref{may8} in Eq.~\eqref{may5}, one may write:
\begin{equation}\label{may9}
\left( -\omega^2 \,  -i \gamma \, \omega+\omega^2_0\right)  \vec R_0 = - \frac{e}{m_0}\, \vec E_0, 
\end{equation}
and hence: 
\begin{equation}\label{may10}
\vec R_0=  \frac{-e\, \vec E_0}{m_e \left( -\omega^2 \,  -i \gamma \, \omega+\omega^2_0\right)}\, , 
\end{equation}
where $\omega$ is the frequency of the electric field. The \emph{electric polarization vector} $\vec P$ is given by \cite{jackson}:
\begin{equation}\label{P2}
\vec P= n_0\, p(t), 
\end{equation}
where $p(t)$ denotes the \emph{dipole moment} per unit volume and $n_0$ is the number of effective electrons per unit volume. One can write:
\begin{equation}\label{P3}
n_0 p(t)= -n_0 \, q\, \mathbf r_i =  \frac{n_0 e^2 \, \vec E_0}{m_e \left[ \omega^2_0-\left( \omega^2 \,  +i \gamma \, \omega\right) \right]}. 
\end{equation}
The \emph{displacement} vector $\vec D$ throughout the volume of the gas is written as \cite{garrity}:
\begin{equation}\label{D0}
\vec D= \vec E+4\pi \, \vec P. 
\end{equation}
Assuming the substance is isotropic, the relation between polarization vector and electric field is given by:
\begin{equation}\label{0-E}
\vec P= -\frac{\chi_e(\omega)}{4\pi }\, \vec E, 
\end{equation}
where $\chi_e(\omega)$ denotes the \emph{electrical susceptibility} of a dielectric material. Using Eq.~\eqref{0-E}, one may write Eq.~\eqref{D0} as:
\begin{equation}\label{D2}
\vec D= \frac{1}{4\pi}\, \varepsilon(\omega)\, \vec E, 
\end{equation}
where we put:
\begin{equation}\label{D3}
\varepsilon(\omega)=1+ \chi_e(\omega), 
\end{equation}
denoting \emph{dielectric function} for a materiel. Substituting Eqs.~\eqref{P3} and \eqref{D3} in Eq.~\eqref{D2}, one finds the dielectric function for a material as:
\begin{equation}\label{1-28}
\varepsilon (\omega) = 1+ \frac{\omega_p^2}{\omega_0^2 - \omega^2 - i\gamma \omega},
\end{equation}
where $\omega_p$ denotes the \emph{bulk plasma frequency} also known as \emph{plasmon} frequency, and is given by:
\begin{equation}\label{1-29}
\omega_p^2= \frac{4\pi\, n_0\, e^2}{m_e}, 
\end{equation}
in \emph{Gaussian's} unit.  The \emph{Drude model} for the dielectric function of metals is obtained in the assumption that the free electrons in metals are not bound to any atoms and are not subject to restoring forces. This implies the spring constant $k$ to be zero and therefore $\omega_0=0$ \cite{Bo,fox}. Using this in Eq.~\eqref{1-28} leads to:
\begin{equation}\label{1-30}
\varepsilon (\omega) =1- \frac{\omega_p^2}{\omega(\omega +i\gamma)}.
\end{equation}
In the absence of damping forces, the free-electron-gas dielectric function reduces to the well-known expression:
\begin{equation}\label{0-eps1}
\varepsilon(\omega)= 1- \frac{\omega^2_p}{\omega^2}.
\end{equation}
In the generalized Drude model one may assume $\gamma$ is frequency-independent and that it is complex valued for $\varepsilon$ to undergo \emph{Kramers Kroing} dispersion relation, in which the real part stays constant with respect to plasma energy, while the imaginary part could be neglected. Thus the simple Drude model stays valid throughout all the steps up  to plasma energy. 
If one ignores the damping force, then the real dielectric function of a free electron gas could be expressed by Eq.~\eqref{0-eps1}. In order to determine the value of dielectric function for surface plasmon oscillations, we impose the requirement of continuity of electric potential and normal component of the displacement vector across the surface of the metal. This is where the geometric effects of the surface  plays a role in finding the values of dielectric function $\varepsilon(\omega)$ and hence the value of frequency $\omega$ itself. We will discuss these geometric effects in the coming chapters thoroughly. 

Throughout the whole discussion, we took on the assumption that all the interactions occur without a delay, i.e. they are instantaneous. In other words, the system is considered in a way that electrostatic solutions are applicable. This gives us a relatively fair approximation as long as our assumption of having large wavevectors for plasmons (comparing to that of light) is valid. This large-wavevector region is known in the literature the as nonretarted region. This allows us to consider \emph{Poisson} and \emph{Laplace}'s equations instead of wave equation to determine the allowed frequencies. What one might loose under this assumption is thoroughly discussed by Ritchie in \cite{RIT34}.  We devote this chapter to describing our approach regarding the  radiative decay of the surface plasmons of an excited metallic surface which mostly follows the one given in \cite{B}. Furthermore, the geometric effect of the curvatures for such metallic surface has been investigated thoroughly in \cite{PassianCurve,PPRB}. 

\section{Preliminaries}
The electrical size is determined by comparing physical dimension of an electronic structure and the signal wavelength. Let us consider a field which varies at low enough frequency when compared to its wavelength. Therefore, considering a small lapse of time, one could assume that the system remains in an internal equilibrium. In other words, the time delay due to wave propagation from one point to any other point can be ignored if the circuit geometry is relatively small compared to the signal wavelength (which prevents the field distribution to vary significantly). Under these circumstances, quantities like potential and surface current (which will be defined later in this chapter) will not be dependent on position and time \cite{sui}. We start this section with elaborating the mathematical formulation in greater detail.

Let us recall \emph{Maxwell's equations} in their most standard formulations as: 
\begin{align}
\vec \nabla \cdot \vec E &= 4\pi\, \rho,\label{Max2} \\
\vec \nabla \times \vec E&= -\frac{\partial \vec B}{\partial t}, \label{0-Max2} \\
\vec \nabla \cdot \vec B&=0, \label{0-Max3}\\
c^2 \, \vec \nabla \times \vec B&= \mathbf J+ \frac{\partial \vec E}{\partial t}, \label{0-Max4}
\end{align}
where $\vec E$, $\vec B$, $\mathbf J$, and $\rho$ denote \emph{electric} field, \emph{magnetic} field, \emph{current} and \emph{charge density}, respectively. As customary, $c$ indicates the speed of light. %Given any function, here $\Phi$ and vector field $A$, if we set:
%\begin{align}\label{1-1}
%\vec E&= - \left[\frac{1}{c}\, \frac{\partial A}{\partial t}+	\vec \nabla \Phi\right], \\
%\vec B&= \vec \nabla \times A, 
%\end{align}
%as well as:
%\begin{align}\label{1-1E1}
%\vec \nabla \vec B -\frac{\partial \vec B}{\partial t} = \mathbf J, 
%\end{align}
%they satisfy Maxwell's equations \cite{garrity}.  
In the absence of any magnetic field $\vec B$, one could set:
\begin{eqnarray}\label{M1}
&&\vec E= - \left[\frac{1}{c}\, \frac{\partial \mathbf A}{\partial t}+	\vec \nabla \Phi(\mathbf r,t)\right], \label{Max1}
%&&\vec \nabla \cdot \vec E= 4\pi\rho,\label{Max2}
\end{eqnarray}
where $\Phi(\mathbf r,t)$ and $\mathbf A$ are \emph{scalar potential} and \emph{vector potential} field, respectively, and $\mathbf r$ denotes the position vector (see any classical reference on this topic, also \cite{garrity,harris,jackson,SET,Sakurai}). Furthermore, if the wave speed is relatively small compared to $c$, one may ignore the first term in the right-hand side of Eq.~\eqref{Max1}. Under this assumption, we can write Eq.~\eqref{Max1} as: 
\begin{eqnarray}\label{M2}
&&\vec E= - \vec \nabla \Phi(\mathbf r,t).
%&&\vec \nabla \cdot \vec E= 4\pi\rho,\label{Max2}
\end{eqnarray}
By means of Eqs.~\eqref{Max2} and \eqref{M2}, we may assume that the scalar potential $\Phi$ satisfies Poisson's equation:
% changes to:
%\begin{equation}\label{1-1E}
%\vec E= - 	\vec \nabla \Phi(\mathbf r).
%\end{equation}
%Using this in Eq.~\eqref{0-Max2}, one could see: 
%\begin{equation}\label{1-1C}
%\vec \nabla \times E= \vec \nabla \times \big[-\vec \nabla \Phi(\mathbf r)\big]=0, 
%\end{equation}
%which shows the property of electric field being quasi-static, contrary to the time-varying field from Maxwell's equations. Moreover, using Eq.~\eqref{1-1E} in Eq.~ \eqref{Max2}, one could write:
\begin{equation}\label{1-1P2}
\vec \nabla^2 \Phi(\mathbf r,t)= -4\pi \rho. 
\end{equation}
%which relates the quasi-static electrical potential to the charge density at any point of the space. 
%and is known as \emph{Poisson's }equation. 
In a charge-free system, Poisson equation is known as Laplace's equation:
\begin{equation}\label{1-1L}
\vec \nabla^2 \Phi(\mathbf r,t)= 0.
\end{equation}
It is of fundamental importance that under proper boundary conditions the solution to either Poisson or Laplace's equations is unique \cite{med}. In terms of boundary conditions, they are discussed in the next section, for the convenience of the reader and the sake of self consistency.

\section{General Approach} 

It is well-known that the retarded potentials are due to time-varying charge densities which are associated with surface plasmons \cite{Bohm}. 
%If these potentials are evaluated at a distance $r$ from the source, one notes that they are out of phase with the oscillating charge whose amount is given by $\sigma = -\omega \, r /c, $ where $\omega$ is the frequency of the oscillation. Electromagnetic signals traveling with finite speed is the cause of this phase-lag. However, if the propagation time, $r/c$ is assumed to be small comparing with the period of oscillation, $\tau= 2\pi / \omega$, then one could ignore the phase-lag as it becomes negligible, hence non-retarded limit. 
In the non-retarded limit, the instantaneous potentials are calculated assuming the source is a static charge density. 
%One may write $\sigma \equiv  -2\pi r /\lambda, $ with $\lambda$ the wavelength of light at frequency $\omega$. 
Retardation for the fields can be ignored near the surface of a particle if the dimensions of the particle are much less than the wavelength of light at frequency $\omega$. 
%Let us put the above explanations with the mathematical aspect. More precisely, time-varying fields are the solution of wave equation, given as:
%\begin{equation}\label{wave}
%\frac{1}{c^2}\vec \nabla ^2 _t \Phi -\vec \nabla ^2 \Phi = F(x,t),
%\end{equation}
%with constant $c$ denoting wave speed and $F(x,t)$ representing the force term. Assuming the dimensions of the particle are much less than the wavelength of light at frequency $\omega$ is equivalent of letting the speed of light approach infinity and the time-dependency in Eq.~\eqref{wave} can be ignored. Lastly, in the absence of any external force, one gets Eq.~\eqref{1-1L}
%\begin{equation}\label{lap}
%\vec \nabla ^2 \Phi = 0, 
%\end{equation}
%which shows why the solution of Laplace equation could be taken as non-retarded limit. 
\par
\begin{figure}[H]
	\centering
	\includegraphics[width=2.5in]{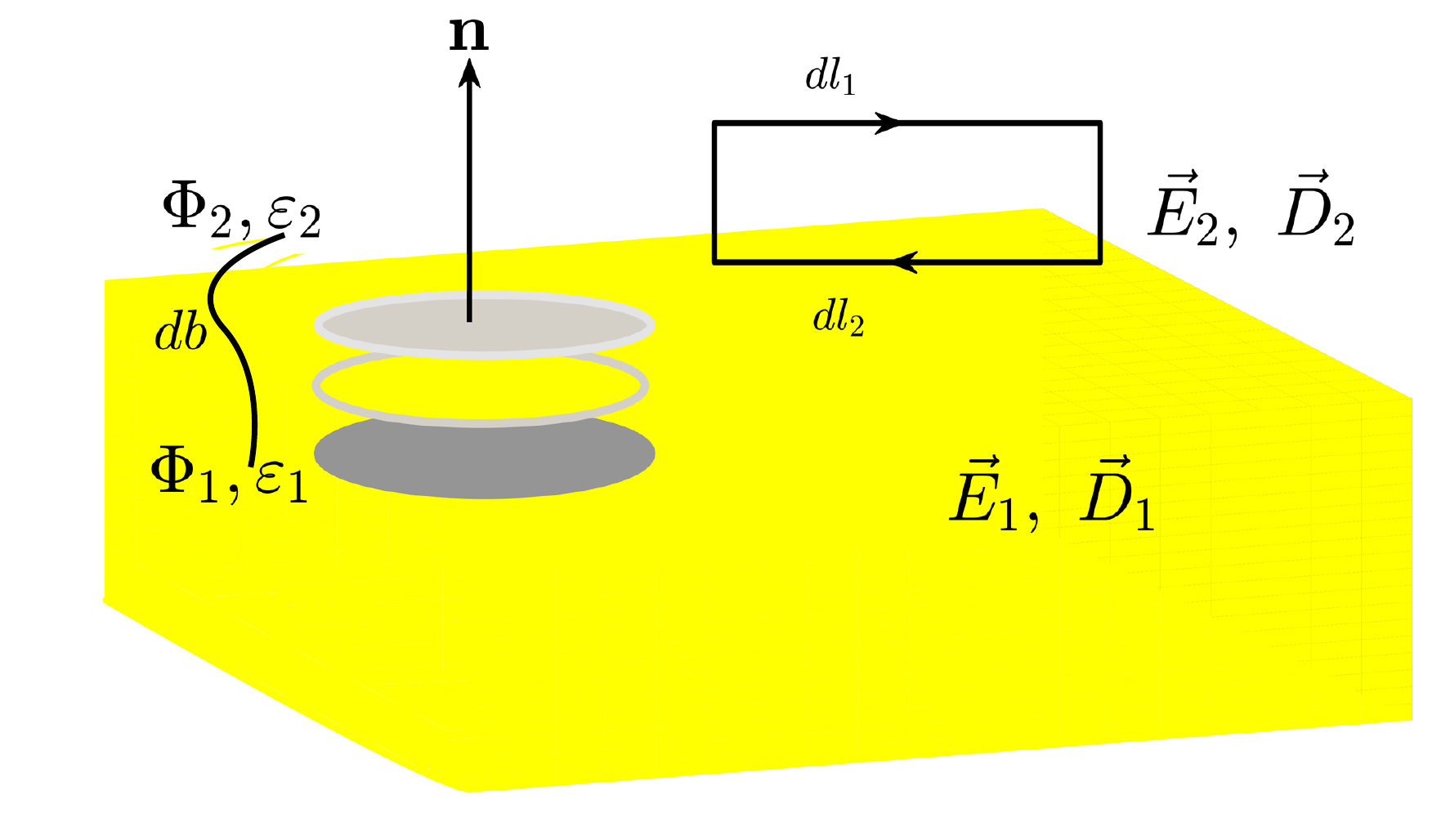}\\	
	\caption[A tiny pillbox of volume $\mathcal V$ at the interface between two]{A tiny pillbox of volume $\mathcal V$ at the interface between two media with dielectric functions $\varepsilon_1$ and  $\varepsilon_2$. Vector $\mathbf n$ is the normal vector to the surface. }
	\label{FBD}
\end{figure}

\begin{figure}[H]
	\centering
	\includegraphics[width=2.5in]{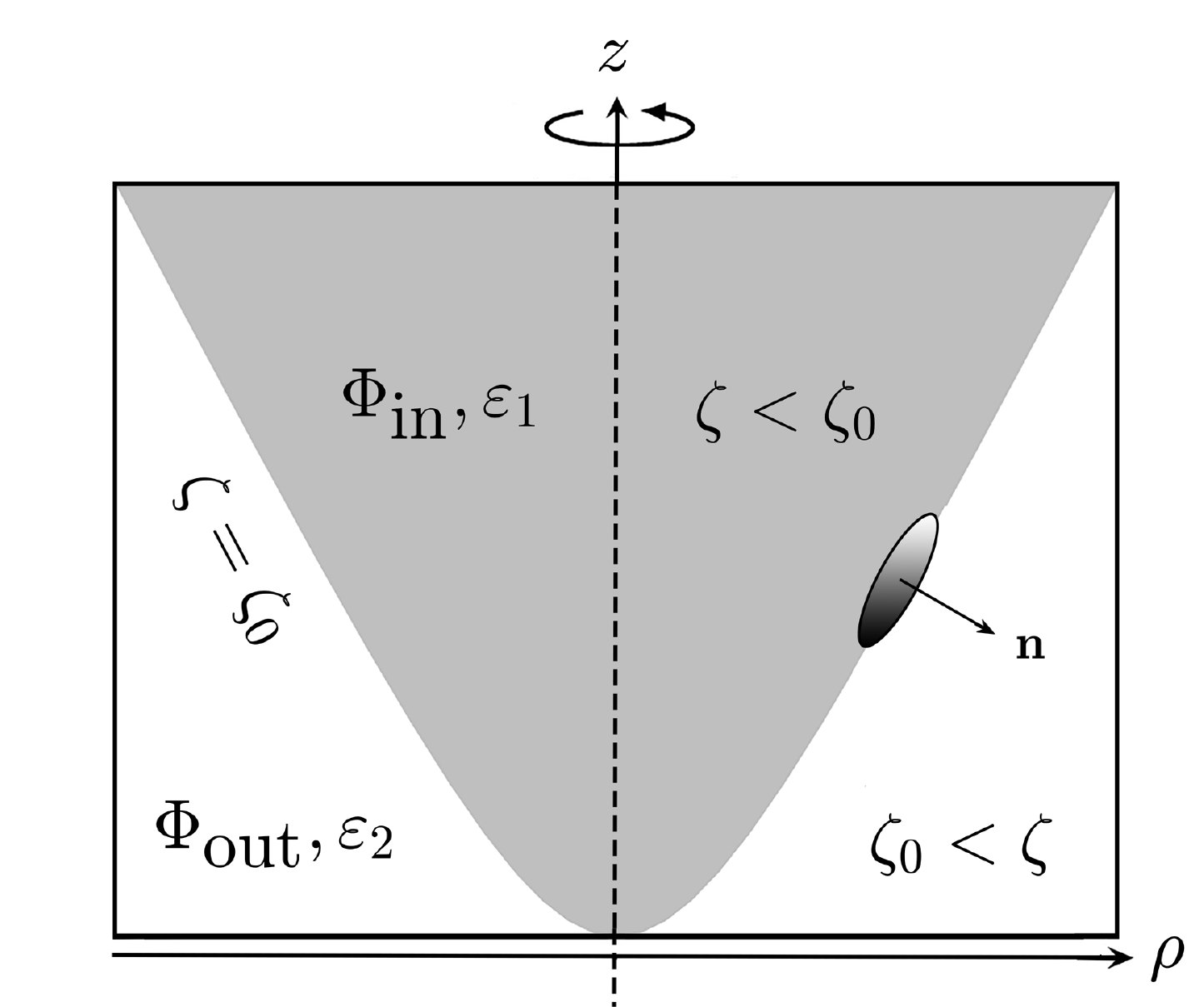}\\	
	\caption[The solid surface of revolution region defined by fixing $\zeta=\zeta_0$, ]{The solid surface of revolution region defined by fixing $\zeta=\zeta_0$, along with inside and outside scalar potentials $\Phi_{\text{in}}$ and $\Phi_{\text{out}}$, respectively. The $\varepsilon_1$ and $\varepsilon_2$ denote the dielectric constant for the inside and outside regions. $\mathbf n$ denotes the normal unit vector. } 
	\label{B}
\end{figure}

 We start by considering a pillbox with volume $\mathcal V$ at the interface of two media with dielectric function $\varepsilon_1$ and $\varepsilon_2$, such that the areas filled with each material are the same (see Fig.~\ref{FBD}). Maxwell's equations imply the continuity of the tangential component of the electric field $\vec E$ and the normal component of displacement vector $\vec D$. We determine these boundary conditions using \emph{Gauss} and \emph{Stoke's} laws \cite{gauss}. 
%\noindent The potential associated with surface plasmons in the non-retarded limit satisfies Laplace equation along with {Dirichlet} and {Neumann} boundary conditions, which require the specific values of $\Phi$ and its normal derivative at each point on the boundary, respectively.
% In terms of 
In view of  Eq.~\eqref{D2}, for any external sources in vacuum with $\varepsilon(\omega)=1$, using Gauss's law for a continuous charge density $\rho$, we can write: 
\begin{equation}\label{Pi1}
\oint_{\partial \Pi} \vec D \cdot \mathbf n \, d\mathcal A= \int_{\Pi} \rho \,d\mathcal V.
\end{equation}
Utilizing the identity:
\begin{equation}\label{Pi3}
\oint_{\partial \Pi} \vec D \cdot \mathbf n \, d\mathcal A=(\vec D_2-\vec D_1)\cdot \mathbf n\,  \Delta \mathcal A, 
\end{equation}
where $\vec D_1$ and $\vec D_2$ correspond to displacement vectors of the inside and outside regions and $\mathbf { n}$ is the surface unit normal vector. Since the polarization charge is only confined to the surface, we may write:
\begin{equation}\label{Pi2}
\int_{\Pi} \rho \,d\mathcal V= \int_{\partial \Pi} \sigma d\mathcal A= \sigma\, \Delta \mathcal A.
\end{equation}
It follows from Eqs.~\eqref{Pi3} and \eqref{Pi2}:
\begin{equation}\label{B1}
(\vec D_2-\vec D_1)\cdot \mathbf n=\sigma, 
\end{equation}
where $\sigma$ notes the {surface charge} density on the surface boundary $\partial \Pi$. Eq.~\eqref{B1} presents no polarization charge and therefore indicates that the normal component of displacement vector $\vec D$ has a shift equals to surface charge density $\sigma$. This is the first boundary condition. Using Eq.~\eqref{D2}, Eq.~\eqref{B1} may be also written as:
\begin{equation}\label{B2}
(\varepsilon_2 \, \vec E_2-\varepsilon_1\, \vec E_1)\cdot \mathbf n=4\pi\, \sigma. 
\end{equation}
Making use of Eq.~\eqref{M2}, we may also write:
\begin{equation}\label{B3}
\varepsilon_1\nabla \Phi_1 \cdot \mathbf{ n}\,  \Big|_{\partial \Pi}-
\varepsilon_2\nabla \Phi_2 \cdot \mathbf {n}\,  \Big|_{\partial \Pi}= 4\pi\, \sigma, 
\end{equation}
the second boundary condition is widely used to determine the dielectric function and, as a consequence, normal mode frequencies for the given geometry. Using the definition $\oint_C\vec E\cdot dl= 0$ on a closed path $C$, it follows that:
\begin{equation}\label{B4}
\int_{A}^{B} \vec E\cdot dl= - (\Phi_{B}  - \Phi_{A}). 
\end{equation}
In virtue of {Stoke's} theorem, we have:
\begin{equation}\label{B5}
\int_{\partial \Pi} (\vec \nabla\times\vec E) \cdot \mathbf n\,  d\mathcal A= 0.
\end{equation}
Under electrostatic condition, we derive the identity $\vec \nabla \times E=0$. Considering a closed loop as shown in Fig.~\ref{FBD} and applying the line-integral along the loop, we obtain:
\begin{equation}\label{B6}
\oint_C \vec E \cdot dl= \int (\vec E_2- \vec E_1) \cdot dl = 0, 
\end{equation}
having used the fact that $dl_1= -dl_2$. In vector notation, since Eq.~\eqref{B6} holds for any component $dl$, it makes the tangential component of $\vec E$ to be continuous along the boundary. Therefore, 
\begin{equation}\label{B7}
(\vec E_2- \vec E_1)\times \mathbf n=0. 
\end{equation}
By means of Eq.~\eqref{D2}, it follows from Eq.~\eqref{B7} that:
\begin{equation}\label{B8}
\frac{\vec D_1}{\varepsilon_1}= \frac{\vec D_2}{\varepsilon_2}. 
\end{equation}
Moreover, using Eq.~\eqref{M2}, we can obtain:
\begin{equation}\label{B9}
\Phi_{2} -\Phi_{1} = \int (\vec E_2- \vec E_1)\cdot db, 
\end{equation}
where $db$ is shown in Fig.~\ref{FBD}. Once $db\to 0$, the right hand side of Eq.~\eqref{B9} vanishes. Hence at the boundary we have:
\begin{equation}\label{B10}
\Phi_{1} =\Phi_{2}. 
\end{equation}
%\noindent Using two parameters $r$ and $\beta$, such a boundary surface can be parameterized as~\cite{Ency}: 
%\begin{equation}
%\mathbf r= \mathbf r(r,\beta)= r\cos\beta\vec i + r\sin\beta \vec j + f(r) \vec k,
%\end{equation}
%for any explicit equation of a surface of revolution written as $z=f(r),$ where $\mathbf r$ and $r=|\mathbf r|$ represent the usual Cartesian position vector and its length, respectively. 
Given a coordinate system $(\zeta, \xi, \varphi)$  with the scaling factors :
\begin{equation}
h_\zeta= \left| \frac{\partial \mathbf {r}}{\partial \zeta}\right |, \quad
h_\xi= \left| \frac{\partial \mathbf { r}}{\partial \xi}\right |, \quad
h_\varphi= \left| \frac{\partial \mathbf{  r}}{\partial \varphi}\right |, 
\end{equation}
we consider a body $\Pi$ with dielectric function $\varepsilon_1$ immersed in a medium with dielectric function $\varepsilon_2$. Boundary surface $\partial \Pi$, finite or infinite, is described via revolution through an angle $\varphi$ about the $z$--axis resulting in the usual azimuthal symmetry and is defined by fixing one of the coordinates, say $\zeta=\zeta_0$ \cite{Ency}. Furthermore, we denote the inside and outside regions of $\Pi$ by $\zeta<\zeta_0$ and $\zeta>\zeta_0$ as well as their respective potentials as $\Phi_{\text{i}}$ and $\Phi_{\text{o}} $, respectively, noting that: 
\begin{equation}\label{in-out}
\begin{cases}
\nabla^2 \Phi_{\text{i}} = 0, \quad &\zeta<\zeta_0, \\
\nabla^2 \Phi_{\text{o}} = 0, \quad &\zeta_0<\zeta,
\end{cases}
\end{equation}
while at the boundary $\partial \Pi$, %using $\vec D=\varepsilon \vec E$ \cite{Love_Plasma}:
\begin{equation}\label{BD1}
\Phi_{\text{i}}	\, \Big|_{\zeta=\zeta_0}= \Phi_{\text{o}}	\, \Big|_{\zeta=\zeta_0},
\end{equation}
\begin{equation}\label{BD22}
\varepsilon_1 \nabla \Phi_{\text{i}} \cdot \mathbf{ n}\,  \Big|_{\zeta=\zeta_0}=
\varepsilon_2	\nabla \Phi_{\text{o}} \cdot \mathbf {n}\,  \Big|_{\zeta=\zeta_0}.
\end{equation}
Considering the quotient $\varepsilon_1/\varepsilon_2$, we may assume with no loss of generality, $\varepsilon_2=1$ which corresponds the dielectric constant of vacuum, and $\varepsilon_1=\varepsilon$. Thus Neumann boundary condition given in Eq.~\eqref{BD22} can be rewritten as:
\begin{equation}\label{BD2}
\varepsilon \nabla \Phi_{\text{i}} \cdot \mathbf{ n}\,  \Big|_{\zeta=\zeta_0}=
\nabla \Phi_{\text{o}} \cdot \mathbf {n}\,  \Big|_{\zeta=\zeta_0}, 
\end{equation}
An arbitrary surface of revolution along with inside and outside potentials and dielectric constants are illustrated in Fig.~\ref{B}. \\ 

\section{Laplace equation}

Since the non-retarded potential $\Phi$'s inside and outside parts satisfy the Laplace equation (see Eq.~\eqref{in-out}), we consider general form of the Laplacian in a coordinate system $(\zeta, \xi, \varphi)$ :
\begin{multline}\label{0-L}
\vec \nabla^2 \Phi= \frac{1}{h_\zeta\, h_\xi\, h_\varphi}\, \Bigg\{\frac{\partial}{\partial \zeta} \left[ \frac{h_\xi\, h_\varphi}{h_\zeta}\, \frac{\partial \Phi}{\partial \zeta}\right] \\
+ \frac{\partial}{\partial \xi} \left[ \frac{h_\zeta\,  h_\varphi}{h_\xi}\, \frac{\partial \Phi}{\partial \xi}\right]+\frac{\partial}{\partial \varphi} \left[ \frac{h_\zeta\, h_\xi}{h_\varphi}\, \frac{\partial \Phi}{\partial \varphi}\right]
\Bigg \}.
\end{multline}
A coordinate system is called separable if it obeys \emph{Robertson} condition (see \cite{morse}, p.~655). Except the case of a nanoring in which a quasi-separation of variables holds, all considered coordinate systems in this manuscript satisfy Robertson condition and are separable. The case of a nanoring is treated separately in Chapter \ref{ring}. 
%namely, that the scale factors $h_\zeta, h_\xi$ and $h_\varphi$, in the Laplacian given in Eq.~\eqref{0-L}, can be written in terms of functions $f_i$ for $i=1,2,3$ of each component, here $\zeta, \eta$ and $\varphi$, such that $S$ can be written as:
%\emph{Stackel Determinant} $S$ is used to decide in which coordinate system the Laplace equation is separable and is defined by \cite{morse,MSFTH}:
%\begin{equation}
%S= \begin{vmatrix}
%\Phi_{11}(\zeta )& \Phi_{12}(\zeta )&\Phi_{13}(\zeta ) \\ 
%\Phi_{21}(\xi) & 	\Phi_{22}(\xi)& 	\Phi_{23}(\xi)\\ 
%\Phi_{31}(\varphi) & \Phi_{32}(\varphi) & \Phi_{33}(\varphi)
%\end{vmatrix}.
%\end{equation}
%where $\Phi_{ij}(\cdot) $ are some functions of coordinates $\zeta$, $\xi$ and $\varphi$. 
%\begin{equation}\label{0-s}
%S= \frac{h_\zeta\, h_\eta\, h_\varphi}{f_1(\zeta)\, f_2(\eta)\, f_3(\varphi)}.
%\end{equation}
%When is true, one could apply separation of variables method to find the solutions of Laplace equation. \\
Inspired by the azimuthal symmetry assumption and hence the $2\pi$-periodicity of the potential $\Phi$, an ansatz of the form:
%With the assumption of Laplace's equation being separable in given coordinate system, one can assume the ansatz:
\begin{equation}\label{0-p}
\Phi_m= Z(\zeta)\, E (\xi)\, e^{im\varphi}, 
\end{equation}
where $m=0, \pm 1, \pm 2, \dots$ is used. As a consequence, we obtain two \emph{Sturm-Liouville} problems in $\zeta$ and $\xi$ (see \cite{morse} p.~656 and Appendix A, Eqs.~\eqref{0-SL}--\eqref{0-SL2}). Denoting by $F_{mn}(\zeta)$, $\tilde F_{mn}(\zeta)$, and $G_{mn}(\xi)$, $\tilde G_{mn}(\xi)$ the linearly independent pair of solutions to the corresponding obtained Sturm-Liouville problems in $Z(\zeta)$ and $E (\xi)$, where $n$ stands for the second separation constant, 
the general form of eigenfunctions for the Laplacian may be written as:
\begin{equation}
\Phi_{mn}(\zeta, \xi, \varphi)= \begin{cases} F_{mn}(\zeta)\\ \tilde F_{mn}(\zeta) \end{cases}
\times \begin{cases} G_{mn}(\xi)\\ \tilde G_{mn}(\xi) \end{cases} \times e^{im\varphi}.
\end{equation}
For simplicity, we may assume that $n$ takes discrete values, i.e. $n=0, \pm 1, \pm 2, \dots$. However, as we shall see in Chapter \ref{3}, we will also deal with real continuous spectrum with respect to $n$. Utilizing the \emph{superposition} principle, we may write the general solution as:
%\begin{equation}
%\Phi(\zeta, \xi, \varphi)= \sum\limits_{m, n\in \mathbb{Z}} \, \mathcal D_{mn}\, F_{mn}(\zeta)\, G_{mn}(\xi)\, e^{im\varphi}, 
%\end{equation}
%where $\mathcal D_{mn}$ could be determined by boundary conditions. 
%Laplace equation does not contain any time-derivatives which enables us to have  a time-dependence general solution. The time-dependence potential is given by:
\begin{multline}
\Phi(\zeta, \xi, \varphi, t)= \sum\limits_{m,n\in \mathbb{Z}} \, \mathcal D_{mn}(t)\,\begin{cases} F_{mn}(\zeta)\\ \tilde F_{mn}(\zeta) \end{cases}
\times \begin{cases} G_{mn}(\xi)\\ \tilde G_{mn}(\xi) \end{cases} \\\times e^{im\varphi}, 
\end{multline}
where $\mathcal D_{mn}(t)$ is a general Fourier component with respect to time $t$ and they are considered as the non-retarded, time-dependence amplitudes at time $t$. Consequently, the Inside and outside potentials, $\Phi_{\text{i}}$ and $\Phi_{\text{o}}$, could be defined by considering the asymptotic behavior of the eigenfunctions $F_{mn}(\zeta)$, $\tilde F_{mn}(\zeta)$  and $G_{mn}(\xi)$, $ \tilde G_{mn}(\xi)$ as the potential has to satisfy the following conditions:
\begin{eqnarray}\label{NBC}
\begin{cases}
&	\Phi(\mathbf r,t)~\text{is finite at every point of space and for all $t$, }\\
&\lim \limits	_	{r\to\infty}		\Phi(\mathbf r,t)=0, \quad \text{where} \quad r=|\mathbf {r}|. \notag 
\end{cases}
\end{eqnarray}
Without loss of generality, we may assume the sets $ F_{mn}(\zeta)\, G_{mn}(\xi)$ and $ \tilde F_{mn}(\zeta)\, G_{mn}(\xi)$ determine finite potentials for the  inside and the outside of the geometry, respectively. Using Eq.~\eqref{NBC} together with boundary condition given in Eq.~\eqref{BD1}, we may write:
\begin{eqnarray}
\Phi_{\text{i}}(\mathbf r,t)&=&\sum\limits_{m,n\in \mathbb{Z}} \, \mathcal D_{mn}(t)\, F_{mn}(\zeta) \tilde F_{mn}(\zeta_0) 
G_{mn}(\xi)\, \notag \\
&& \qquad \qquad \times  e^{im\varphi}, \qquad \zeta\le \zeta_0, \label{in}\\
\Phi_{\text{o}}(\mathbf r,t)&=&\sum\limits_{m,n\in \mathbb{Z}} \, \mathcal D_{mn}(t)\,   F_{mn}(\zeta_0) \tilde F_{mn}(\zeta) 
G_{mn}(\xi)\,\notag  \\
&& \qquad \qquad \times  e^{im\varphi}, \qquad \zeta_0\le \zeta. \label{out}
\end{eqnarray}
%The elements $ \tilde F_{mn}(\zeta_0)$ and $  F_{mn}(\zeta_0)$ are also chosen based on the boundary condition given in Eq.~ \eqref{BD1}.

\section{Harmonic Oscillator Model}\label{HA}

In this section, we will show that amplitudes $\mathcal D_{mn}(t)$ undergo harmonic oscillator motion. We start by deriving two relations between surface charge density and quasi-static potential using the \emph{total charge} for the whole space. Equating the two relations gives us an equation in amplitudes $\mathcal D_{mn}(t)$ and their second time-derivative $\ddot {\mathcal D}_{mn}(t)$ known as the harmonic oscillator equation of motion, see Eq.~\eqref{0-HA}. The total charge $Q$ is defined by:
\begin{equation}\label{Q1}
Q=\int_{\text{volume}} \rho \, d\mathcal V,
\end{equation} 
where $\rho$ denotes the volume-charge density. Since the polarization charge is only confined to the surface, Eq.~\eqref{Q1} may be written as:
\begin{equation}\label{Q2}
Q= \int_{\text{surface}} \sigma \, d\mathcal A, 
\end{equation} 
where $\sigma$ denotes the surface charge density and $d\mathcal A$ is the surface element. Equating Eq.~\eqref{Q1} with Eq.~\eqref{Q2}, we have:
\begin{equation}\label{Q3}
\int_{\zeta<\zeta_0}^{\zeta>\zeta_0} \int \int \rho \, h_\zeta h_\xi h_\varphi \, d\zeta d\xi d\varphi= 
\int \int_{\zeta =\zeta_0}\sigma  \,  h_\xi h_\varphi \, d\xi d\varphi . 
\end{equation} 
It follows now from Eq.~\eqref{Q3} that
\begin{equation}\label{0-25}
\rho= \frac{\delta (\zeta-\zeta_0)}{h_\zeta} \sigma, 
\end{equation}
where $\delta$ denotes the \emph{Dirac delta} function. In view of Poisson's equation given in Eq.~\eqref{1-1P2}, one finds: 
\begin{equation}\label{0-26}
\vec \nabla^2 \Phi = -\frac{4\pi\, \delta(\zeta-\zeta_0)}{h_\zeta}  \sigma.
\end{equation}
On the other hand, using the electric polarization vector given by \cite{jackson}:
\begin{equation}\label{P1}
\vec P= n_0 (-e) \, \vec u,
\end{equation}
where $\vec u$ denotes the \emph{charge displacement} of the system. Assuming there is no free-charge density on surface, then using Eq.~\eqref{D0}, we can write:
\begin{equation}\label{D1}
\nabla \cdot  \vec D= \nabla \cdot (\vec E + 4\pi \vec P)= 0. 
\end{equation}
It follows from  Eq.~\eqref{Max2}  that $\vec \nabla \cdot  \vec P= -\rho$, which in view of Eq.~\eqref{Q1} and  \emph{Divergence Theorem} gives:
\begin{eqnarray}
Q&=& -\int _{\text{volume}}  \vec \nabla \cdot \vec P\, d\mathcal V\\
&=& -\oint _{\text{surface}}  \vec P\cdot \mathbf{ n} \, d\mathcal A\\
&=& \int _{\text{surface}} \vec P\cdot \mathbf{\vec e}_{\zeta}\big|_{\zeta=\zeta_0} \, d\mathcal A,
\end{eqnarray} 
%from the. This theorem allows us to pass from an integral of any vector filed $F$, over a compact region in space through the following relation: 
%\begin{equation}\label{0-div}
%\int\int\int_{\Omega} \vec \nabla \cdot F\,  d\mathcal V
%= \int\int_{\partial \Omega} F\cdot \mathbf{ n} \, d\mathcal A.
%\end{equation} 
where $\mathbf{ n}=-\mathbf{\vec e}_{\zeta}$. 
Comparing the above equation with Eq.~\eqref{Q2} and using Eq.~\eqref{P1}, one can find:
\begin{equation}\label{sigma0}
\sigma= \vec P \cdot \mathbf{\vec e}_{\zeta}\big| _{\zeta=\zeta_0}= -n_0 e \, \vec u\cdot\,  \mathbf{\vec e}_{\zeta}\big| _{\zeta=\zeta_0}.
\end{equation}
The surface charge due to polarization
is also derived as the motion of free electrons in the metal surface due to a time-harmonic external electric field, following standard descriptions of anomalous dispersion \cite{SET}. We can thus also describe this surface charge as $
\sigma=-n_0\, e\, \vec u. $
The electrostatic force on the electron current per unit volume is $
F=-n_0e\nabla\Phi_{\text {i} }.$
Putting these together through Newton's force law, since  the acceleration is the force per unit mass, we have:
\begin{equation}\label{0-ddotu}
\ddot {\vec  u} =\frac{e}{m_e} \nabla\Phi_{\text {i}}. 
\end{equation}
Differentiating both sides of Eq.~\eqref{sigma0} twice with respect to time gives:
\begin{equation}\label{sigmaddot}
\ddot{\sigma}= -n_0 \, e\,  \ddot{\vec{u}}\cdot \mathbf{\vec e}_{\zeta}\big| _{\zeta=\zeta_0}, 
\end{equation}
which together with Eq.~\eqref{0-ddotu} implies:
\begin{eqnarray}\label{0B34}
\ddot{\sigma}=-\frac{\omega_p^2}{4\pi }\, \mathbf{\vec{e}}_\zeta\, \cdot\vec \nabla\Phi_{\text {i}}|_{\zeta=\zeta_0}.
\end{eqnarray}
Utilizing the identity:
\begin{eqnarray}\label{0-iden}
\mathbf{\vec{e}}_\zeta\, \cdot\vec \nabla= \frac{1}{h_\zeta}\, \frac{\partial }{\partial \zeta}, 
\end{eqnarray}
one may write:
\begin{equation}\label{0-sddot1}
\ddot{\sigma}=-\frac{\omega_p^2}{4\pi } \bigg(
\frac{1}{h_{\zeta}}\, \frac{\partial\Phi_{\text {i}}}{\partial\zeta}
\bigg)\Bigg|_{\zeta=\zeta_0}.
\end{equation}
This equation is also provided in \cite{B} without a detailed proof. Differentiating Eq.~\eqref{0-26} twice with respect to time $t$ gives:
\begin{equation}\label{0-sddot}
\vec \nabla^2 \ddot{\Phi }= -\frac{4\pi\, \delta(\zeta-\zeta_0)}{h_\zeta}  \ddot{\sigma}.
\end{equation} 
By equating Eq.~\eqref{0-sddot1}, in terms of ${\mathcal D}_{mn}(t)$, and Eq.~\eqref{0-sddot}, in terms of second time-derivatives of amplitudes $\ddot 	{\mathcal D}_{mn}(t)$, we obtain the harmonic oscillator equation with the frequencies $\omega_{mn}$ of the form:
\begin{equation}\label{0-HA}
\ddot 	{\mathcal D}_{mn}(t) + \omega_{mn}^2 \, \mathcal D_{mn}(t)=0.
\end{equation}
% these frequencies are equal to those which are obtained  by solving second boundary condition given in Eq.~\eqref{BD2} and using Drude model given in Eq.~\eqref{0-eps1} (see \cite{zw}). 
The importance of Eq.~\eqref{0-HA} is the fact that the  (allowed) resonant  values of the dielectric function $\varepsilon$ can also 
be independently calculated from this transcendental equation obtained by imposing the quasistatic boundary conditions at the bounding surfaces of the domain within which the scalar electric field satisfies the Laplace equation \cite{zw}.\\

\section{Classical Energy}

Once the scalar potential and charge density are known, we shall be able to calculate the \textit{total energy} of the system. In classical mechanics, the total energy $E$, also known as \textit{total mechanical energy}, of a position dependent force, in our case surface plasmons, is given by:
\begin{equation}\label{0-39}
E=V+T, 
\end{equation}
where $V$ and $T$ denote \textit{potential} and \textit{kinetic} energy, respectively. This is nothing more than an expression of the conservation of mechanical energy principle \cite{hasbun,buhler}. We shall start by calculating potential energy. \\
The potential energy is given by the volume integral \cite{B}:
\begin{equation}\label{0-35}
V= \frac12 \int_{\zeta\le \zeta_0} \rho\,  \Phi_{\text{i}}\, d\mathcal V, 
\end{equation}
where $\mathcal V$ is the volume of the geometry of interest. Using Eq.~\eqref{0-25}, one could write:
\begin{equation}\label{0-36}
V= \frac12 \int_{\zeta \le \zeta_0} \frac{\delta (\zeta-\zeta_0)}{h_\zeta}\,  \sigma \, \Phi _{\text{i}}\, d\mathcal V.
\end{equation}
Since inside and outside scalar potential $\Phi_{\text{i}}$ and $\Phi_{\text{o}}$ coincide on the surface, namely the first boundary condition given in Eq.~\eqref{BD1}, it is a direct computation using any of the relations of inside or outside scalar potentials. Utilizing the  $\delta$ function, we can write:
\begin{equation}\label{0-37}
V =\frac12 \int_{\text{surface}}
\sigma\,   {\Phi_{\text{i}}}\bigg|_{\zeta=\zeta_0}  \, 
d\mathcal A.
\end{equation}
Using the identity $d\mathcal A= h_{\xi }h_{\varphi} \, d\xi  d\varphi$, we find:
\begin{equation}\label{0-V}
V =\frac12 \int\int  
\sigma  {\Phi_{\text{i}}}\bigg|_{\zeta=\zeta_0}  
h_{\xi }h_{\varphi} \, d\xi d\varphi. 
\end{equation}
The kinetic energy $T$ of the oscillations is given by the volume integral \cite{B}:
\begin{equation}\label{0-T}
T= \frac{m_e n_0}{2}
\int_{ \textrm{volume}}
\dot{\vec{u}}  \cdot{\dot{\vec{u}}}\, d\mathcal V.
\end{equation}
Depending on whether time-dependent amplitudes obey pure harmonic oscillator or not, we may take different steps in order to find kinetic energy. 
In all the cases investigated here, amplitudes satisfy the relation given in Eq.~\eqref{0-HA} but the case for nano-ring. We proceed by assuming the general case and we defer the discussion for nano-ring to Chapter \ref{ring}. 
We may use the relation given in Eq.~\eqref{0-ddotu} to write: 
\begin{equation}\label{0-udd1}
\ddot {\vec  u} =\frac{e}{m_e} \, \vec \nabla\Phi_{\text {i}}= \frac{e}{m_e}\, \vec \nabla\ \sum\limits_{mn} \mathcal D_{mn}(t) \Psi_{mn}, 
\end{equation}
where we put: 
\begin{equation}\label{0-psi}
\Phi_{\text {i}}= \sum\limits_{mn}\Phi_{mn}= \sum\limits_{mn} \mathcal D_{mn}(t)\Psi_{mn}. 
\end{equation}
Now, using Eq.~\eqref{0-HA}, we may write:
\begin{equation}
\mathcal D_{mn}(t)= -\frac{\ddot {\mathcal D}_{mn}(t)}{\omega_{mn}^2}. 
\end{equation}
Putting this in Eq.~\eqref{0-psi}, we could write:
\begin{equation}\label{0-psi1}
\Phi_{\text {i}}= \sum\limits_{mn}\ddot {\Phi}_{mn}=-\sum\limits_{mn} \frac{\ddot {\mathcal D}_{mn}(t)}{\omega_{mn}^2}\, \Psi_{mn}. 
\end{equation}
Replacing this is Eq.~\eqref{0-udd1}, we obtain:
\begin{equation}
\ddot {\vec  u} =- \frac{e}{m_e}\, \vec \nabla\sum\limits_{mn} \frac{\ddot {\mathcal D}_{mn}(t)}{\omega_{mn}^2}\, \Psi_{mn}. 
\end{equation}
Integrating both sides of the above equation with respect to time $t$ gives:
\begin{equation}\label{0-udot}
\dot{\vec  u}=- \frac{e}{m_e}\, \vec \nabla\sum\limits_{mn} \frac{\dot{\mathcal D}_{mn}(t)}{\omega_{mn}^2}\, \Psi_{mn}= - \frac{e}{m_e}\, \vec \nabla\sum\limits_{mn} \frac{	\dot{\Phi}_{mn}}{\omega_{mn}^2}. 
\end{equation}
Using the latter expression for $\dot{\vec u}$ in Eq.~\eqref{0-T}, we may write: 
\begin{equation}\label{0-T1}
T= \frac{ e^2\, n_0}{2\, m_e}
\int_{ \textrm{volume}}
\left( \vec \nabla\sum\limits_{mn} \frac{	\dot{\Phi}_{mn}}{\omega_{mn}^2}\right) 
\cdot  \left( \vec \nabla\sum\limits_{m'n'} \frac{	\overline{\dot{\Phi}_{m'n'}}}{\omega_{m'n'}^2}\right) \, d\mathcal V, 
\end{equation}
where $\overline{\dot{\Phi}_{mn}}$ denotes the conjugate of ${\dot{\Phi}}_{mn}$. The fact that scalar potential is real-valued allows us to replace it with its complex conjugate. Here, \emph{Fubini's Theorem} \cite{conway} allows us to change the order of the summation and integration, assuming the integral is bounded. Using above argument as well as plasmon-bulk frequency relation given in Eq.~\eqref{1-29}, Eq.~\eqref{0-T1} could be rewritten as:
\begin{multline}\label{0-T2}
T= \frac{\omega_p^2}{8\pi}\, \sum\limits_{mn}\, \sum\limits_{m'n'}\frac{1}{\omega_{mn}^2\, \omega_{m'n'}^2}\\ \times 
\int_{ \textrm{volume}}
\left( \vec \nabla 	\dot{\Phi}_{mn}\right) 
\cdot \left( \vec \nabla 	\overline{\dot{\Phi}_{m'n'}}\right) \, d\mathcal V. 
\end{multline}
Since Laplace equation is independent from time $t$, it is easy to see that if $\Phi_{\text {i}}$ satisfies the Laplace equation, so does its time-derivative  $\dot\Phi_{\text{i}}$. We can use the following general identity: 
\begin{equation}\label{1-49}
\vec \nabla
\dot\Phi \cdot
\vec \nabla \dot\Phi=
\vec \nabla\cdot
\left ( \dot\Phi
\vec \nabla
\dot\Phi \right ),
\end{equation}
to write the kinetic energy as: 
\begin{multline}\label{1-50}
T= \frac{\omega_p^2}{8\pi}\, \sum\limits_{mn}\, \sum\limits_{m'n'}\frac{1}{\omega_{mn}^2\, \omega_{m'n'}^2} \\ \times 
\int_{ \textrm{volume}}
\vec \nabla\cdot \, 
\left( 
\dot\Phi_{mn}\, 
\vec \nabla
\overline{\dot\Phi_{m'n'}}
\right) 
\, d\mathcal V. 
\end{multline}
The volume integral in Eq.~\eqref{1-50}, using Divergence Theorem could be transformed to a surface integral through the following relation:
\begin{multline}\label{1-51}
T= \frac{\omega_p^2}{8\pi}\, \sum\limits_{mn}\, \sum\limits_{m'n'}
\frac{1}{\omega_{mn}^2\, \omega_{m'n'}^2} \\ \times 
\int_{ \textrm{surface}}
\left( 
\dot\Phi_{mn}\, 
\vec \nabla \overline{\dot\Phi_{m'n}}
\right) 
\bigg |_{\zeta=\zeta_0}\, 
\cdot \, 
\mathbf{\vec e}_{\zeta}
\, d\mathcal A.
\end{multline}
Using the identity given in Eq.~\eqref{0-iden}, we find the final expression for kinetic energy as: 
\begin{multline}\label{Sp-K}
T=\frac{\omega_p^2}{8\pi }   
\sum\limits_{mn} 
\sum\limits_{m'n'}
\frac{1}{\omega_{mn}^2\, \omega_{m'n'}^2} \\ \times 
\int\, \int 
\dot \Phi_{mn} \, \frac{\partial\overline{\dot \Phi_{m'n'}}}{\partial \zeta}\, 
\frac{h_\xi\, h_\varphi}{h_\zeta}\, 
d\xi d\varphi .	
\end{multline}
From Eqs.~\eqref{0-V} and \eqref{Sp-K}, the classical energy is obtained by using Eq.~\eqref{0-39}. Depending on amplitudes being real or complex valued, they could be written in the following form:
\begin{equation}\label{0-D}
\mathcal D_{mn}(t)= \frac{\mathbf{\alpha_{mn}}}{2\omega_{mn}} \big (d_{mn}\pm d^*_{mn}\big ) , 
\end{equation}
with time-dependent operators $d_{mn}$ directly proportional to $e^{i\omega_{mn}t}$, where $d^*_{mn}$ is its respective complex conjugate. The purely imaginary case could be assumed by multiplying the real case of Eq.~\eqref{0-D}  by $i$. For the simplicity of the discussion, we may consider the case for which amplitudes are real-valued. The goal is now to determine the values of $\mathbf{\alpha_{mn}}$. First find the time-derivative of amplitudes $\mathcal D_{mn}(t)$ using Eq.~\eqref{0-D}, i.e.: 
\begin{equation}\label{0-Dd}
\dot{	\mathcal D}_{mn}(t)=i\, \frac{\mathbf{\alpha_{mn}}}{2} \big (d_{mn}- d^*_{mn}\big ). 
\end{equation}
Using Eqs.~\eqref{0-D} and \eqref{0-Dd} to write the total energy $E$ in terms of $d_{mn}$ and $d^*_{mn}$ and comparing the new expression with \emph{Hamiltonian} operator for the quantum mechanical harmonic oscillator given as \cite{pines}:
\begin{equation}\label{0-H}
H\equiv \sum\limits_{mn} \frac{\hbar \omega_{mn}}{2}\, \left({\hat d}^\dagger_{mn}\, \hat d_{mn}+ \hat d_{mn}\, \hat d^\dagger _{mn}\right), 
\end{equation}
where $\hbar$ is the \emph{Planck's} constant . The following substitution allows us to find the values for $\mathbf {\delta_{mn}}$:
\begin{equation}\label{0-S}
\left( d_{mn}\, , \, d^*_{mn}\right)  \rightarrow \left( {\hat d}_{mn}\, , \, \hat d^\dagger _{mn}\right). 
\end{equation}
The operators $\left( {\hat d}_{mn}\, , \, \hat d^\dagger _{mn}\right)$ are interpreted as the non-commuting \emph{boson} creation and annihilation operators, respectively. 
Calculations of the interactions between surface plasmons and other excitations are often time complicated, yet have been simplified by quantizing the fields (second quantization) \cite{Ritchie}. The total classic energy was calculated to help us finding coefficients in Hamiltonian operator. In the next section, we will use the Hamiltonian operator to calculate matrix element and probability amplitudes for interactions.

\section{Quantum Field \& Second Quantization }\label{QF}

\par
Among the many quantum numbers that distinguish particles, we can count their spin \cite{zeid, maggiore, Folland}. There are two main families of particles: \emph{bosons}, with integer spin and \emph{fermions}, with half-integer spin. Photons, gluons and vector bosons are examples of bosonic particles and electrons, neutrinos, and quarks are classified as fermionic particles.

There have been two main approaches to quantum field theory and on how to study the interaction between particles and surfaces. One approach is due to Feynman: His celebrated path-integral approach based on \emph{Feynman's diagram}. Two detailed accounts can be found in \cite{zeid}. This approach does not rely on operators and the theory behind the operator in \emph{Hilbert} spaces. Another approach, by means of operator theory, can be traced back to Heisenberg, Born, Jordan, Dirac, Pauli, and von Neumann. This method, more formal than the previous one, is more difficult to apply to practical computations, due to the nonlinearities of the interactions. In \cite{zeid}, the crucial aspect of a quantum field is that it can be treated as a system with infinitely many quantum particles that can be created and annihilated. The main purpose of second quantization is to describe quantum particles, either bosons or fermions, in terms of operators. The number of particles is assumed to be infinite. We first set our full system as the combined surface plasmon and photon field. Then we consider the ground state or zero-population for both plasmon and photon as:
\begin{equation}
|0\rangle.
\end{equation}
A state consisting of precisely $N$ plasmon particles is shown by the notation:
\begin{equation}
|\text{Plasmon} \rangle\equiv | \nu _{m_1n_1} \, \cdots \, \nu _{m_Nn_N}\, \cdots \rangle, 
\end{equation}
where $\nu _{mn}$'s are the number of plamons in the state $(mn)$. Similarly, the state consisting of exactly $\tilde N$ photons can be represented by:
\begin{equation}
|\text{Photon} \rangle\equiv | \nu _{s_1 q_1} \, \cdots \, \nu _{s_{\tilde N} q_{\tilde N}}\, \cdots \rangle, 
\end{equation}
where $\nu _{ s q}$'s show the number of photons with wavevector $s$ and polarization $q=1,2$. Polarization vectors are chosen based on the fact that they are perpendicular to each other and to the direction of travel. It is customary to represent them by s-polarization and p-polarization vectors. We shall discuss it in more details in Chapter \ref{3}.
Our main purpose  concerns the investigation of interacting quantum fields in rigorous mathematical terms. We shall use the fact that the operators describing photons only act on and hence change the photon state. The same situation holds for the plasmon operators while acting over the full system. The states being orthonormal results in the Kronecker delta function as:
\begin{equation}
\langle \cdots \nu_{n_k} \, \cdots \, \nu_{n_1} \big| \nu_{n'_1}\, \cdots \, \nu_{n'_k}	\cdots  \rangle= \delta_{n_1n'_1}\, \cdots \, \delta_{n_kn'_k}\, \cdots, 
\end{equation}
for some arbitrary states $1, \cdots, k$. Considering these facts, we are now able to write a general state for the combined plasmon and photon field in the form:
\begin{multline}
|\text{Photon} \rangle\, |\text{Plasmon} \rangle\equiv\,\\
 | \nu _{s_1 q_1} \, \cdots \, \nu _{s_{\tilde N} q_{\tilde N}}\rangle\,  | \nu _{m_1n_1} \, \cdots \, \nu _{m_Nn_N}\, \cdots\, \rangle. 
\end{multline}
Next, we shall discuss how creation and annihilation operators, $\hat d_{mn}$ and $\hat d^\dagger_{mn}$ acting on an arbitrary state, including ground state of the field. A prototype of commutation relations for these operators when act on bosonic particles is given in \cite{zeid, maggiore}:
\begin{equation}\label{0-r1}
\langle \, 0\, |\, \hat d _{mn}\, \hat d^\dagger _{m'n'}\, |\, 0\rangle=\delta_{mm'}\, \delta_{nn'}, 
\end{equation}
for all plasmon states $mn$ and $m'n'$ and 
\begin{equation}\label{0-r2}
\langle 0\, |\, \hat a_{s'q'}\, \hat a^*_{sq}\, |\, 0\rangle=\delta(s-s')\, \delta_{qq'}.
\end{equation}
with $\hat a$ and $\hat a^*$, creation and annihilation operators for photon and for all photon states $sq$ and $s'q'$.

\section{Quantization of the Electromagnetic field }\label{1-f}

%Let us consider the Electromagnetic field on a large finite box $\mathcal B$ that vanishes at the ends. 
In view of Maxwell's equations (recall Eqs.~\eqref{Max2}--\eqref{0-Max4}), both electric and magnetic fields can be expressed in terms of vector potential $\mathbf A$ as:
\begin{align}
\vec E&= - 	\vec \nabla \Phi(\mathbf r,t)-\frac{\partial \mathbf A}{\partial t}, \\
\vec B&= \vec \nabla \times \mathbf A. 
\end{align}
In case of absence of any sources of the field, it is possible to choose the \emph{Coulomb gauge} condition as:
\begin{equation}\label{1-1A}
\vec \nabla\cdot  \mathbf A = 0.
\end{equation}
This condition is also known as \emph{transversality} condition since the vector potential under this condition is purely transverse and could be expressed as \cite{harris,hatfield}:
\begin{multline}\label{A}
{\mathbf A}=\sum\limits_{q=1,2}
\int \frac{1}{(2\pi)^3}
\sqrt{\frac{\hbar c^2 }{\omega_s}}\,
\mathbf {\hat e}_q\, 
\Big[a_{sq}(t)\, e^{i\,  s\cdot \mathbf{r}}\\+  
a^*_{sq}(t)\, e^{-i\, s\cdot \mathbf{r}}\, \Big] d^3s\,,
\end{multline}
where $\hbar\omega_s$ is the photon energy, the polarization vector $\mathbf {\hat e}_q$ is perpendicular to $s$ for both values of polarization on index $q=1,2$ and $ a_{sq}(t)$ and its conjugate $ a^*_{sq}(t)$ are the photon operators, such that:
\begin{equation}\label{A1}
a_{sq}(t)= a_{sq}(0)\, e^{-i\omega_st}, 
\end{equation}
therefore:
\begin{equation}\label{A2}
\dot a_{sq}(t)=-i\omega_s\, a_{sq}(t), 
\end{equation}
which is considered as the equations for motion of the field for all $s$. To write the photon field as a sum with discrete momentum eigenstates as opposed to the continuous representation, we have to make some specific choices. We consider our electromagnetic field to be confined to a volume $\mathcal{V},$ which is normally taken to be represented by a cube over which we impose periodic boundary conditions. 
Since the electromagnetic energy confined to this volume is independent of the shape of the volume \cite{Bennett, Belinsky, Folland}, we take as our quantization volume as a box $\mathcal B$, with side $L$. 
The transition to the discrete sum then follows the \emph{lattice strategy} in quantum electrodynamics~\cite{zeid} which we summarize as follows:
\begin{enumerate}
	\item Periodic free field: Classical free field is represented in terms of finite Fourier coefficients. The first step is to confine the photons and electrons within the above-mentioned box $\mathcal B$.
	\item We impose periodic boundary conditions by assuming that fields have the period $L$ with respect to the position of particles. Observe that this new condition on boundary by no means contradicts the boundary conditions we first defined in Eqs.~\eqref{BD1} and \eqref{BD2}.
	\item We introduce a momentum space which yields finite Fourier series. 
	\item The Fourier coefficients of the finite Fourier series are replaced by creation and annihilation operators we introduced in this section. This is the bridge from classical mechanics to quantum mechanics. 
	\item Using finite number of creation and annihilation operators, the matrix elements could be formed. 
	\item Finally, we let the finite box $\mathcal B$ tend to infinity, in other words, we undertake the limit of $L\to \infty$. We let  the period $L$ become infinite.  
\end{enumerate} 
In practice, we perform the transition:
\begin{eqnarray}\label{0- FV}
\frac{1}{(2\pi)^{3/2}}	\int d^3 s\rightarrow\frac{1}{\sqrt{\mathcal{V}}}\sum\limits_{ s},
\end{eqnarray} 
to write the discretized vector potential as:
\begin{eqnarray}\label{0-A3}
\mathbf A=
\sum\limits_{s}
\sum\limits_{q=1,2}
\sqrt{\frac{\hbar c^2 }{\mathcal V \omega_s}}
\mathbf{\hat e}_q
\big (	\hat a_{sq}	\,e^{i\mathbf{s\cdot r}}	+\hat a^{\dagger}_{sq}   \,e^{-i\mathbf{s\cdot r}}	\big ). 
\end{eqnarray}
where the photon operators have been replaced by $\hat a_{sq}\, , \hat a^\dagger_{sq}$, annihilation and creation operators for photon. Following Eq.~\eqref{0-r2}, we replace Dirac delta function $\delta(s-s')$ by the discrete \emph{Kronecker delta} function $\delta_{ss'}$ and obtain new  the commutation relations as:
\begin{equation}\label{1-cm}
[\hat a_{sq},\hat a^{\dagger}_{sq}]=\delta_{qq'}\delta_{ss'}.
\end{equation} 
Given the explicit form of vector potential $\mathbf A$, the goal is to calculate the \emph{probability amplitude} and \emph{interaction Hamiltonian}.\\

\noindent \textbf{Probability amplitude and matrix elements}

The probability amplitude that a surface plasmon in a given state, defined by $m',n'$ will interact with a photon, described by $s',q'$ is expressed as:
\begin{equation}
P(m'n'\to s'q')=\Big |\langle 0|\nu_{s'q'}\, H_{\text{int}}\, \nu^*_{m'n'}|0\rangle\Big |^2, 
\end{equation}
where $H_{\text{int}}$ denotes the interaction Hamiltonian. We shall call: 
\begin{equation}
\mathbf{\mathcal M}=\langle 0|\nu_{s'q'}\, H_{\text{int}}\, \nu^*_{m'n'}|0\rangle\, , 
\end{equation}
as the interaction \emph{matrix element}. We could define various matrix elements of the interaction Hamiltonian based on the type of interaction.  A collection of matrix elements are given in \cite{B} and we list them here:
\begin{enumerate}
	\item Direct scattering (elastic, including the Thomson limit):
	\begin{equation}\label{Mds}
	\mathbf{\mathcal M}_{ds}=
	\left\langle 0\left| \hat a_{q_{f}}(s_{f})\, 
	H^{(0)}_{RP}~\hat a^*_{q_i}(s_{i})\right| 0\right\rangle,
	\end{equation}
	where index $f$ shows final state while index $i$ indicates initial state and $H^{(0)}_{RP}$ denotes the direct scattering interaction Hamiltonian, given by: 
	\begin{equation}\label{Hds}
	H^{(0)}_{RP}=
	\frac{n_0e^2}{2mc^2}
	\int_{\zeta\le \zeta_0}\mathbf A\cdot\mathbf A\, d\mathcal V.
	\end{equation}
	where $\zeta_0$ is the shape parameter. This Hamiltonian is also known as the zeroth-term of Hamiltonian. 
	%%%%%%%%%%%%%%%%%%%%%%%%%%%%%%%%%%%%%%%%%%%%%%%%%%%%%%%%%%%%%%%%%%%%%%%%%

	\item Emission (radiative decay of surface plasmons):
	\begin{equation}\label{Mem}
	\mathbf{\mathcal M}_{em}= 
	\left\langle 0\left| \hat a_{q_{f}}(s_{f})~
	H^{(1)}_{RP}~\hat d^{\dagger}_{m_in_i}\, \right| 0\right\rangle,
	\end{equation}
	where
	\begin{equation}\label{Hem}
	H^{(1)}_{RP}=\frac{1}{c}\int \mathbf J\cdot\mathbf A\, d\mathcal V.
	\end{equation}
	The term $H^{(1)}_{RP}$ represents the interaction of one photon and one plasmon, which can be used to predicct the creation of a plasmon by a photon or the decay of a plasmon into a photon.

	%%%%%%%%%%%%%%%%%%%%%%%%%%%%%%%%%%%%%%%%%%%%%%%%%%%%%%%%%%%%%%%%%%%%%%%%%
	\item Absorption:
	\begin{equation}\label{Mabs}
	\mathbf{\mathcal  M}_{abs}= \left\langle 0\left| \hat d_{m_fn_f}\, H^{(1)}_{RP}\, \hat a^*_{q_{i}}(s_{i})\right| 0\right\rangle, 
	\end{equation}
	which is the Hermitian dual to Eq.~\eqref{Mem}, since it represents the inverse process. 
	%%%%%%%%%%%%%%%%%%%%%%%%%%%%%%%%%%%%%%%%%%%%%%%%%%%%%%%%%%%%%%%%%%%%%%%%%
	\item Total elastic scattering:
	Here $\kappa$ is the plasmon damping factor and the matrix element is obtained as shown below:
	\begin{multline}\label{Mtes}
	\mathbf{\mathcal M}_{tes}= 	\mathbf{\mathcal M}_{ds}+\sum_{mn}\frac{1}{\hbar}\Bigg[\frac{\mathbf{\mathcal M}_{em}	\mathbf{\mathcal M}_{abs}}{\omega_{s_{i}}-\omega_{km}+(i\kappa/2)}\\-\frac{\mathbf{\mathcal M}^*_{em}	\mathbf{\mathcal M}^*_{abs}}{\omega_{s_{f}}+\omega_{km}-(i\kappa/2)} \Bigg] .
	\end{multline}
	%%%%%%%%%%%%%%%%%%%%%%%%%%%%%%%%%%%%%%%%%%%%%%%%%%%%%%%%%%%%%%%%%%%%%%%%%
	\item Inelastic scattering:
	\begin{equation}\label{Mc}
	\mathbf{\mathcal M}_{c}= \left\langle 0\left| \hat a_{q_{f}}(s_{f}) \hat d_{m_fn_f}\, H^{(2)}_{RP}\, \hat a^*_{q_{i}}(s_{i})\right| 0\right\rangle,
	\end{equation}
	where 
	\begin{equation}\label{Hc}
	H^{(2)}_{RP}=\frac{e^2}{2mc^2}\int \mathbf{\hat n}\, \mathbf A\cdot\mathbf A\, d\mathcal V.
	\end{equation}
	This term involves two photons and one plasmon and can describe, for example, the inelastic scattering of a photon in creating a plasmon. 
\end{enumerate}
\section{ Interaction Hamiltonian}

Here, to describe the plasmon-photon  interaction, we resort to the hydrodynamical formulation of the electron gas by Crowell and Ritchie~\cite{Crowell} as:
\begin{eqnarray}\label{0-FPC}
&&\vec\nabla \frac{\partial}{\partial t} \Psi ({\bf r}, t)= -\frac em \vec\nabla \Phi({\bf r}, t)+ \frac{\xi^2}{n_0}\vec\nabla n({\bf r}, t), \\
&&\vec\nabla^2 \Phi({\bf r}, t)=4\pi e \, n({\bf r}, t), \\
&&\vec\nabla^2\Psi({\bf r}, t) =\frac{\partial}{\partial t} n({\bf r}, t)/n_0,  
\end{eqnarray}
where $\Phi(\bf r, t)$, $\Psi(\bf r, t)$ and $n(\bf r, t)$ are the electric potential, velocity potential, and electronic density, respectively, in the electron gas, while $n_0$ is the electronic density in the undisturbed state of the electron gas and $\xi$ is the propagation speed of the disturbance through electron gas. By linearizing these equations, employing perturbation theory, 
Ritchie obtained the first order interaction Hamiltonian given as:
\begin{equation}\label{0-Int}
{H}_{\text{int}}=
\frac{1}{c}
\int\mathbf{J\cdot A}\, d\Omega, 
\end{equation}
where $\mathbf J$ denotes the \emph{current density}. This interaction has been used previously to describe the emission of  photons via plasmon decay on finite surfaces of an oblate spheroid modeling  silver nanoparticles evaporated on a dielectric substrate~\cite{Little_Paper}. Here, we will apply this Hamiltonian to modeling the creation of a plasmon on the surface  by a photon or the decay of a plasmon and emission of a photon~\cite{B}. This application requires the explicit determination of the current density operator. Let us first consider the current given by:
\begin{equation}\label{0-J}
\mathbf{J}=-n_e e \, \dot{\vec u}.
\end{equation}
We define:
\begin{equation}\label{0-psidot}
\dot \Psi =- \frac{e }{m_e}\, \sum\limits_{mn} \frac{	\dot{\Phi}_{mn}}{\omega_{mn}^2}, 
\end{equation}
in which we make the replacement of $\dot{D}_{mn}(t)$ with its equivalent given in Eq.~\eqref{0-Dd}, and use Eq. ~\eqref{0-udot} to write:
\begin{equation}\label{0-J1}
\mathbf{J}= n_e \, e \, \vec \nabla\dot \Psi. 
\end{equation}
Thus one can write: 
\begin{equation}
H_{\text{int}}= -\frac{n_0e}{c}\int(\vec \nabla\dot{\Psi})\cdot \mathbf{A} \, d\mathcal V.
\end{equation}
Using the vector identity: 
\begin{equation}
(\vec \nabla\dot{\Psi})\cdot \mathbf A=\vec \nabla\cdot
(\dot{\Psi}	\mathbf A)-\Psi(\vec \nabla\cdot \mathbf A).
\end{equation}
Imposing the Coulomb gauge condition given in Eq.~\eqref{1-1A}, and the divergence theorem, recalling that the current only exists on the surface of the body, we can find: 
\begin{equation}
H_{\text{int}} =  -\frac{n_0e}{c}\int\vec \nabla\cdot
\left(\dot\Psi \mathbf A \right)\, d \mathcal V=-\frac{n_0e^2}{c}
\int_{\zeta=\zeta_0}
\left(\dot \Psi  \mathbf A \right)
\, d\mathcal A,  
\end{equation}
which gives the interaction Hamiltonian as: 
\begin{eqnarray}\label{0-IntH}
H_{\text{int}}  =   -\frac{n_0\, e}{c}
\int\, \int 
\left(	\dot\Psi \mathbf A	\cdot	\mathbf{\hat e}_{\zeta}	\right)\, 
h_\xi	h_\varphi \,	d\xi 	d\varphi. 
\end{eqnarray}
Once $H_{\text{em}}  $ is known, can calculate the emission matrix element given in Eq.~\eqref{Mem}. 
Using Eqs.~\eqref{0-A3} and \eqref{0-psidot}, we write:
\begin{multline}\label{0-IntH1}
H_{\text{int}}  =   -\frac{n_0\, e}{c}
\int\, \int 
\Bigg\{\bigg[	- \frac{e }{m_e}\, \sum\limits_{mn} \frac{	\dot{\Phi}_{mn}}{\omega_{mn}^2} \bigg]\\
\times \bigg[\sum\limits_{s}
\sum\limits_{q=1,2}
\sqrt{\frac{\hbar c^2 }{\mathcal V \omega_s}}
\Big (	\hat a_{sq}	e^{i\mathbf{s\cdot r}}	\\
+\hat a^{\dagger}_{sq}   e^{-i\mathbf{s\cdot r}}	\Big )
\times \left( 	\mathbf{\hat e}_q\cdot	\mathbf{\hat e}_{\zeta}	\right) \bigg] \Bigg\}\, 
h_\xi	h_\varphi \,	d\xi 	d\varphi. 
\end{multline}
Using Fubini's Theorem in Eq.~\eqref{0-IntH1}, we obtain:
\begin{multline}\label{0-IntH2}
H_{\text{int}}  =   \frac{n_0\, e^2}{m_e} \\
\times \sum\limits_{mn}\, \sum\limits_{s}
\sum\limits_{q=1,2} \sqrt{\frac{\hbar  }{\mathcal V \omega_s}}
 \int\, \int 
\Bigg\{\bigg(	  \frac{	\dot{\Phi}_{mn}}{\omega_{mn}^2} \bigg)\\
\times \bigg[		
\left (	\hat a_{sq}	\, e^{i\mathbf{s\cdot r}}	+\hat a^{\dagger}_{sq}   \, e^{-i\mathbf{s\cdot r}}	\right )
\times \left( 	\mathbf{\hat e}_q\cdot	\mathbf{\hat e}_{\zeta}	\right) \bigg] \Bigg\}\, 
h_\xi	h_\varphi \,	d\xi 	d\varphi. 
\end{multline}
We use this equation in the relevant transition matrix element for photon emission, given by Eq.~\eqref{Mem} where $\hat a_{q_fs_f}$ and $\hat d^\dagger_{m_i n_i}$ describe  the final photon state and the initial plasmon state, respectively. It is noteworthy to emphasize again that photon operators act only on the photon state and plasmon operators act only on the plasmon state \cite{Folland, JRTPE, zeid, maggiore, dick}. The question of which particular operators, either creation or annihilation, give rise to a non vanishing matrix element depends entirely on initial and final states. The following  commutation relation is our main reference to calculate the probability amplitude and therefore, the matrix element: 
\begin{equation}\label{0-R}
\left\langle 
0\left| a_{q_f s_f}\, \Big[ \hat a^\dagger _{qs}\, \hat d_{mn}\, \Big] \hat d^\dagger_{m_i n_i} \right|  	0
\right\rangle = \delta _{ s s_f}\, \delta_{q q_f}\, \delta_{mm_i}\, \delta_{nn_i}. 
\end{equation}
Using all these commutation relations, we could discard the summations in Eq.~\eqref{0-IntH2} and can proceed to calculate the matrix element. Having calculated the emission matrix element, we have all the ingredients to discuss and calculate the radiative decay in the following section. 

\section{Radiative Decay Rate}\label{rad}

In the study of the phenomenon of radiation in the quantum domain, the starting point is to talk about vector potential $\mathbf A$ \cite{Sakurai}. Imposing guage condition given in Eq.~\eqref{1-1A}, the fields, either electric or magnetic, which are derived from vector potentials are called \emph{radiation} fields. This term is often used to describe, improperly, the vector potential itself.  \emph{Fermi's} formalism based on emission interaction Hamiltonian, derived in the previous section, is called the radiation or Coulomb gauge method. A theory once quantized, provides transparent details about the process of bosonic particles, here photons, being emitted, absorbed and scattered. Let us now specify our discussion with respect to emission process. We consider a solid angle element $d\Omega$ through which a photon is emitted. The number of allowed states in an energy interval can be written as: 
\begin{equation}\label{0-o1}
\rho_{d\Omega}= \frac{\mathcal V\omega_{s}^2}{(2\pi)^3}\frac{d\Omega}{\hbar c^3},
\end{equation}
with $s$ pointing into the solid angle $d\Omega$, This equation also is known as the the density of a single photon \cite{Sakurai}. Fermi's Golden rule \cite{jackson} gives the transition probability in the form:
\begin{eqnarray}\label{0-g1}
\gamma_{fi}=\frac{2\pi}{\hbar}\, |\mathcal M_{em}|^2 \rho_f,
\end{eqnarray}
where $ \mathcal M_{em}$ represents the matrix element given in Eq.~\eqref{Mem} and $\rho_f$ is the density of final states. The total radiative decay rate, $\gamma_{mn}$, is obtained by summing over final photon states as:
\begin{eqnarray}\label{0-g2}
\gamma_{mn}=\sum_{s}\sum_{q=1,2}\gamma_{fi}.
\end{eqnarray}
If we let the quantization volume, $\mathcal V$, become large enough so that the sum over $ s$ changes into an integral, then we could write:
\begin{eqnarray}\label{0-g3}
\gamma_{mn}=\sum_{q=1,2} \int \gamma_{fi}\, d\mathcal V.
\end{eqnarray}
We could now write, using Eq.~\eqref{0-o1}:
\begin{equation}
\frac{\partial\gamma_{mn}}{\partial \Omega}=\sum_{q=1,2}\frac{\mathcal V}{(2\pi)^3} 
\int\gamma_{fi}\frac{\omega_{\vec s}^2}{\hbar c^3}\,  d\omega_{s}. 
\end{equation}
Lastly, using Eq.~\eqref{0-g1}, we find:
\begin{equation}\label{0-dr}
\frac{d\gamma_{mn}}{d \Omega}=
\sum_{q=1,2}\frac{\mathcal V}{(2\pi)^3} 
\bigg [\frac{2\pi}{\hbar^2}~|\mathcal M^q_{em}|^2 \bigg ]_{\omega_s=\omega_{mn}}\frac{\omega_{mn}^2}{ c^3}, 
\end{equation}
which gives the radiative decay rate per solid angle for two polarization vectors $q=1,2$.
Eq. ~\eqref{0-dr}, as one can easily see, becomes independent from the quantization volume $\mathcal V$, also known as normalized volume. In calculating Eq.~\eqref{0-dr}, we also require to expand the plane wave $
e^{\pm i\, s\mathbf{\cdot r}},$
using the transition expressions from the given coordinate system, here $(\zeta, \xi, \varphi)$, to Cartesian coordinates. Often time, the resulting wave functions are difficult to calculate. Depending on the case, we either avoid direct calculations of wave functions using rotating and scaling the wave vector $s$ with algebraic substitutions, or we used numerical calculations to compute integrals due to the absence of analytical solutions.

In the next chapter, we apply the whole procedure given in the current chapter to two infinite geometries as solid paraboloid and hyperboloid and one finite geometry, prolate spheroid. The results for these two cases have been provided in \cite{Bag2018}.

%In the next chapter, we apply the whole procedure given in the current chapter to two known cases as solid sphere and solid cylinder. The results for these two cases have been provided in several references, mainly \cite{B, Crowell, BURMISTROVA}. However, the detailed calculations  presented here, to the best of our knowledge, can not be found in the literature. 

%\bibliography{references}
%\bibliographystyle{ieeetr}
\end{document}